\documentclass{aa}
\usepackage{graphicx}
\begin{document}
   \title{Highly accurate calculation of rotating neutron stars}

   \author{M. Ansorg, A. Kleinw\"achter and R. Meinel
          }

   \offprints{M. Ansorg,\\ \email{ansorg@tpi.uni-jena.de}}

   \institute{Theoretisch--Physikalisches Institut, University of Jena,
              Max--Wien--Platz 1, D--07743 Jena, Germany\\
              }

   \date{Received / accepted }

   \abstract{
   A new spectral code for constructing general--relativistic models of rapidly
   rotating stars with an unprecedented accuracy is presented.
   As a first application,
   we reexamine uniformly rotating homogeneous stars and compare our results with 
   those obtained by several previous codes. Moreover, representative relativistic 
   examples corresponding to highly flattened rotating bodies are given.
   \keywords{stars: rotation ---
                stars: neutron ---
                gravitation ---
                relativity ---
                methods: numerical
               }
               }
   \maketitle
%
\section{Introduction}

The study of relativistic, axisymmetric and stationary, uniformly
rotating perfect fluid bodies is motivated by extraordinarily 
compact astrophysical objects like neutron stars. Several
numerical codes have been developed
in order to calculate the structure and the gravitational field of
these bodies (Bonazzola \& Schneider \cite{bs}, Wilson \cite{wi},
Butterworth \& Ipser \cite{butips1, butips2}, Friedman et al. \cite{fip1,fip2},
Lattimer et al. \cite{latt}, Neugebauer \& Herold \cite{nh}, Herold \&
Neugebauer \cite{hn}, Komatsu et al. \cite{keh1, keh2}, Eriguchi et al.
\cite{ehn}, Stergioulas \& Friedman \cite{ster}, Bonazzola et al. \cite{bgsm};
for reviews see Friedman \cite{fri} and Stergioulas \cite{sterg}).
While they obtain an accuracy of up to 5 digits for sufficiently smooth
equations of state,
these methods yield fewer than 4 digits in the case of
stiff equations of state (e.g., for constant density),
which is due to particular Gibbs phenomena
at the star's surface. In order to avoid these Gibbs
phenomena, Bonazzola
et al. (\cite{bgm}) used a multi--domain spectral method
with which they were able to achieve an accuracy of 12 digits
for the Maclaurin sequence of homogeneous Newtonian bodies.

In this Letter we introduce a new numerical code, which is based
on a multi--domain spectral method for representing all metric
functions. We intend to use this method to investigate neutron 
stars with realistic equations of state. In particular, our
multi--domain method lends itself  to considering several layers
inside the star, which are characterized by different equations of
state. As we will outline below, we obtain a hitherto unobtainable accuracy
which permits its application even in limiting cases such as the
mass--shedding limit. Moreover, we are able to study
extremely flattened, homogeneous Einsteinian bodies. Such bodies
were the subject of interesting
conjectures made by Bardeen (\cite{bardeen}).

As a first application of our method, we reexamine a particular
example of a uniformly rotating homogeneous star that was used by
Nozawa et al. (\cite{nozawa}) to compare three different codes.
We give our
results, which possess a substantially higher accuracy. Moreover,
we discuss representative, relativistic
examples corresponding to highly flattened rotating bodies.

In what follows, units are used in which the velocity of light as
well as Newton's constant of gravitation are equal to 1.
\section{Metric tensor and field equations}

The line element for an axisymmetric, stationary, uniformly
rotating perfect fluid body assumes in Lewis--Papapetrou coordinates
$(\rho,\zeta,\varphi,t)$ the following form:
\[ds^2=e^{-2U}[e^{2k}(d\rho^2+d\zeta^2)+W^2 d\varphi^2]
         -e^{2U}(dt+ad\varphi)^2\,.\]
To define the coordinates $(\rho,\zeta)$ uniquely, we require that
the metric coefficients and their first derivatives be continous
at the surface of the body.

In the vacuum region, there emerge three field equations of second order to
determine the potentials $U$, $a$ and $W$. The function $k$ follows from the
other potentials by a line integral\footnote{In the highly
relativistic regime, we use $e^{2U}$, $a e^{2U}$ and $k-U$ instead of
$U$, $a$ and $k$ in order to avoid problems in the presence of ergoregions.}.

In the interior of the body we use the metric functions valid in the comoving 
frame of reference. Here, the only new coordinate is
$\varphi'=\varphi-\Omega t$, where $\Omega$ is the angular velocity of the body.
The corresponding line element also assumes the above form with potentials
$U'$, $a'$, $k'$ and $W'$, which are given by
\[e^{2U'}=e^{2U}[(1+\Omega a)^2-\Omega^2 W^2 e^{-4U}],\]
\[(1-\Omega a')e^{2U'} = (1+\Omega a)e^{2U},\]
\[k'-U'=k-U \quad \mbox{and} \quad W'=W.\]
Since in the comoving frame the energy--momentum tensor reads
    \[T^{ik}=(\mu+p)u^iu^k+pg^{ik}\,,\quad
    u^k=e^{-U'}\delta_4^k\,,\]
where $\mu$ is the mass--energy density and $p$ the pressure,
the field equations assume a particularly simple form, see, e.g.,
Kramer et al. (\cite{kra}), equations (19.35a--c).
For a given equation of state, $p=p(\mu) $ or $\mu=\mu(p)$, the relativistic
Euler equations $T^{ik}_{\quad;k}=0$ yield
\[e^{U'}\exp\left[\int_0^p\frac{dp'}{\mu(p')+p'}\right]
=e^{V_0}=\mbox{const.}\]
Hence, pressure and density can be expressed in terms of $U'$. At the surface
$B$ of the star, the pressure vanishes which leads to a constant
surface potential $U'=V_0$.  
In particular,
for homogeneous stars ($\mu=\mbox{const.}$)
we obtain \[p=\mu(e^{V_0-U'}-1).\] 
Taking formula (19.35d)\footnote{Note that the bar over $\zeta$ in equation
(19.35d) of
Kramer et al. (\cite{kra}) is a misprint.}
and the condition (19.37) of Kramer et al. (\cite{kra})
we may express $k'$ by the
potentials $a'$, $U'$ and $W$ via a line integral. Thus again a
system of three field equations emerges.

All potentials satisfy regularity conditions at infinity and along the axis of
rotation ($\rho=0$) and possess moreover reflectional symmetry with respect to the
plane $\zeta=0$ (see Meinel \& Neugebauer \cite{meineu}).
\section{The numerical scheme}

The numerical scheme to solve the field equations with respect to boundary and
transition conditions is based on a multi--domain spectral method.
After imposing reflectional symmetry, the set of all
relevant $(\rho,\zeta)$--values,
$\{(\rho,\zeta)\mbox{:}\quad\rho\geq 0,\,\zeta\geq0\}$, is
divided into several subregions for physical reasons. In the simplest
case, we only take two subregions, the interior and the exterior of the body.
However, if we consider several layers inside the star, which are characterized 
by different equations of state, we will be forced to allow for more than two
regions. In this Letter, we will restrict ourselves to only two subregions.

Each of these subregions is mapped onto the square $I^2=[0,1]^2$. In order to do
this we introduce a function $y_B$ defined on the Interval $I=[0,1]$ with
\[y_B(0)=1\,,\quad y_B(1)=0, \]
which describes the surface of the body by
\[B=\{(\rho,\zeta)\mbox{:}\quad\rho^2=r_e^2t,\,\zeta^2=r_p^2y_B(t)\,,
\quad 0\leq t\leq 1\},\]
where $r_e$ and $r_p$ are the equatorial and polar coordinate radii of
the body respectively. As a particular example for the mapping in question, we
used for the interior the transformation
\[\rho^2=r_e^2st\,,\quad \zeta^2=r_p^2sy_B(t)\,,\quad (s,t)\in I^2\]
and for the exterior 
\[\rho^2=\frac{r_e^2 t}{s^2}\,,\quad 
\zeta^2=\frac{r_p^2 y_B(t)}{s^2}\,,\quad (s,t)\in I^2.\]
In this manner, the axis $\rho=0$ and the plane $\zeta=0$ are mapped to the
coordinate boundaries $t=0$ and $t=1$ respectively. Furthermore, the surface
$B$ of the body is characterized by $s=1$. For the interior and exterior 
transformation, the point $s=0$ corresponds to the origin and 
to infinity respectively.

We assume all potentials to be smooth functions on $I^2$ such that we may
approximate them well by two--dimensional Chebyshev--expansions with
respect to the coordinates $s$ and $t$. In the same manner, we represent the
unknown boundary function $y_B$ as well as the boundary values $a'_B$ and $W_B$ 
of the potentials $a'$ and $W$ in terms of (one--dimensional)
Chebyshev--polynomials with respect to the coordinate $t$. If these three
one--dimensional functions were given, we would have to solve a particular interior
and exterior boundary value problem\footnote{The boundary values are completed 
by $U'=V_0.$} of the respective field equations. However, we have to deal with
a free boundary value problem, where these three functions are not known, but have
to be determined such that the normal derivatives of the potentials $U'$, $a'$
and $W$ behave continuously at the surface $B$ \footnote{It is a consequence of
the field equations that $k'$ is then also differentiable.}. Taking only a
finite number of Chebyshev--coefficients into account for the interior
potentials $U'$, $a'$ and $W$, the exterior potentials $U$,
$a$ and $W$ and the surface quantities $y_B$, $a'_B$ and $W_B$, our numerical
scheme consists in determining all unknown Chebyshev--coefficients by
satisfying the interior and exterior field equations at a number of grid points
in $I^2$ and, moreover, requiring the above transition conditions at the
surface. The total number of unknown coefficients equals the total number of
equations. The system is solved by a Newton--Raphson method, where the initial
guess for the entire solution in the case of constant density
can be taken from the analytical Newtonian Maclaurin solution.
\section{Results}
\subsection{Comparison with previous codes}
Nozawa et al. (\cite{nozawa}) compared three different codes for various
choices of the
equation of state. We take one particular example in which for constant
density  they prescribed
the normalized central pressure $\bar{p}_c=p_c/\mu=1$ and the
ratio $r_p/r_e=7\times
10^{-1}$. The results of this comparison are given in table 11 of
Nozawa et al. (\cite{nozawa}). Here we
calculate the same quantities with an accuracy of 12 digits and list them in
column 2 of Table 1\footnote{Note that there are misprints in
Nozawa et. al. (\cite{nozawa}) concerning the
formulae for  $Z_p$, $Z_{eq}^f$ and $Z_{eq}^b$: In equation
(26), $-2\nu_p$ has to be replaced by $-\nu_p$, and in equations (27) and (28),
$(\nu_e-\beta_e)/2$ by $(\beta_e-\nu_e)$.}. Columns
3--5 refer to the codes by Komatsu et al. (\cite{keh1}, \cite{keh2}) and Eriguchi
et al. (\cite{ehn})
[abbreviated by KEH(OR)],
Stergioulas and Friedman (\cite{ster}) [abbreviated by KEH(SF)] and Bonazzola
et al. (\cite{bgsm}) [abbreviated by BGSM] and give the relative error of
the quantity in question, i.e. for example,
$|\bar{M}[\mbox{KEH(OR)}]-\bar{M}|/\bar{M}\approx0.023$. The fact that
$r_p/r_e$ was not exactly $0.7$ in the BGSM calculation does not affect
the comparison substantially.
In Nozawa et al. (\cite{nozawa}),
the general--relativistic virial identities GRV2 and GRV3 (derived by Bonazzola \&
Gourgoulhon \cite{bonaz} and Gourgoulhon \& Bonazzola \cite{gourg})
were calculated to check the
accuracy of the numerical solution. For our result, this check yields $1.6\times
10^{-13}$ for GRV2 and $4.4\times 10^{-13}$ for GRV3. Note that we used 23
Chebyshev--polynomials for each dimension in this calculation.
As an  additional test of accuracy,
we calculated the angular momentum and the gravitational
mass in two different ways: (i) from the asymptotic
behaviour of the metric and (ii) by means of integrals over
the matter distribution [cf. Bardeen
\& Wagoner (\cite{bw2}), equations (II.24), (II.26) and (II.23), (II.25)
respectively].
We get a relative deviation of
$1.7\times 10^{-14}$ for the mass and $6.2\times 10^{-14}$ for the angular momentum.
           \begin{table}
           \begin{tabular}{lllll}\hline
           \multicolumn{2}{c}{}&\multicolumn{1}{c}{(1)}&
           \multicolumn{1}{c}{(2)}&\multicolumn{1}{c}{(3)}\\ \hline
           $\bar{p}_c$& 1& & & \\
           $r_p/r_e$&$0.7$ & & &0.11\% \\
           $\bar{\Omega}$&$1.41170848318$& 1.1\%&0.32\% &0.97\%\\
           $\bar{M}$&$0.135798178809$& 2.3\%&0.19\% &0.86\%\\
           $\bar{M}_0$&$0.186338658186$& 0.17\%&0.32\% &1.4\%\\
           $\bar{R}_{circ}$&$0.345476187602$& 0.098\%&0.053\% &0.27\%\\
           $\bar{J}$&$0.0140585992949$& 1.6\%&0.045\% &2.3\%\\
           $Z_p$&$1.70735395213$& 6.1\%&0.013\% &2.1\%\\
           $Z_{eq}^f$&\hspace{-2.6mm}$-0.162534082217$& 1.7\%&1.9\% &4.4\%\\
          $Z_{eq}^b$&$11.3539142587$& 17\%&0.10\% &8.1\%\\ \hline\vspace*{2mm}
         \end{tabular}
         \caption{Results for the constant density
         model with $\bar{p}_c=1$, $r_p/r_e=7\times
          10^{-1}$, and relative errors of the codes
          (1): KEH(OR), (2): KEH(SF) and (3): BGSM.
           $\bar{\Omega}=\Omega/\mu^{1/2}$, $\bar{M}=M \mu^{1/2}$,
           $\bar{M}_0={M}_0 \mu^{1/2}$,
           $\bar{R}_{circ}={R}_{circ}\,\mu^{1/2}$ and $\bar{J}=J\mu$ are
           normalized values of the angular velocity $\Omega$, gravitational
           mass $M$, rest mass $M_0$, equatorial circumferential radius
           ${R}_{circ}$ and angular momentum $J$. $Z_p$ is the
           polar redshift, $Z_{eq}^f$ and $Z_{eq}^b$ are equatorial redshifts
           for photons emitted in the forward and backward direction.}
         \end{table}
\subsection{Highly flattened rotating bodies}
Chandrasekhar (\cite{chandra}) has shown that the post--Newtonian Maclaurin
spheroids become singular at the eccentricity
$\epsilon =\epsilon_1 = 0.98522$\dots,
the point of the first axisymmetric secular instability.
Bardeen (\cite{bardeen}) confirmed this result and discussed the possibility of two
Newtonian axisymmetric sequences bifurcating from the Maclaurin spheroid at
$\epsilon=\epsilon_1$, a first one that should finally evolve towards the
Dyson--Wong rings (Dyson \cite{dys1, dys2}, Wong \cite{wong}, Kowalewsky
\cite{kow}, Poincar\'e \cite{poinc1, poinc2, poinc3}; see also Lichtenstein
\cite{licht}) and a second one, which he called the `central bulge configuration',
that should end in a mass--shedding limit. Eriguchi \& Sugimoto (\cite{eri1})
indeed found the first sequence, and called it the `one--ring sequence'. The
properties of the bifurcation have been analyzed by Christodoulou et al.
(\cite{christo}).

By means of a Newtonian version of our code we were able to
find the second sequence as well. It appears as a continuation of the one--ring
sequence across the Maclaurin sequence and is characterized by a (bi--convex)
`lens shape' of the solutions.

The point $\epsilon=\epsilon_1$ of the Maclaurin sequence corresponds to the
value $R=R_1=0.27320$\dots\, of the rotation parameter
$R=\bar{J}^2/\bar{M}_0^{10/3}$. Bardeen (\cite{bardeen}) speculated that there
should be a gap in the $R$--values of relativistic solutions around $R_1$ and
that the relativistic solutions should show some properties of the Newtonian
`central bulge' or Dyson ring solutions when approaching this gap from the
sphere end or from the disc end of the Maclaurin sequence.

           \begin{table}
           \begin{tabular}{llll}\hline
           \multicolumn{1}{c}{}& \multicolumn{1}{c}{(a)}&\multicolumn{1}{c}{(b)}&
           \multicolumn{1}{c}{(c)}\\ \hline
           $\bar{p}_c$&0.002 &0.004 &0.003 \\
           $r_p/r_e$&0.2 &0.1 &0.1 \\
           $\bar{\Omega}$&1.089468\,e+00 &1.004798\,e+00 & 9.305304\,e$-$01\\
           $\bar{M}$&8.314919\,e$-$04 &1.099261\,e$-$02 &5.603298\,e$-$03\\
           $\bar{M}_0$&8.353271\,e$-$04 &1.124879\,e$-$02  &5.677891\,e$-$03\\
           $\bar{R}_{circ}$&1.028186\,e$-$01 &2.474602\,e$-$01 &2.249448\,e$-$01\\
           $\bar{J}$&3.663135\,e$-$06 &3.093789\,e$-$04 &1.072577\,e$-$04\\
           $Z_p$&1.585397\,e$-$02 &9.132868\,e$-$02  &5.381392\,e$-$02\\
           $Z_{eq}^f$&\hspace{-2.6mm}$-$9.890231\,e$-$02 &
            \hspace{-2.6mm}$-$1.912324\,e$-$01  &
            \hspace{-2.6mm}$-$1.720749\,e$-$01\\
           $Z_{eq}^b$&1.308359\,e$-$01 &3.827610\,e$-$01  &2.825350\,e$-$01\\ \hline
           $R$&0.2444465 &0.3001205 &0.3522896\\ \hline
           $GRV2$&1.1\,e$-$09 &3.1\,e$-$11 &1.9\,e$-$08  \\
           $GRV3$&1.6\,e$-$09 &3.2\,e$-$11 &2.3\,e$-$08 \\ \hline \vspace*{2mm}
         \end{tabular}
         \caption{Three highly flattened, constant density models.
         The corresponding surface shapes can be found in Fig.~1.
         In these calculations we used 23 Chebyshev--polynomials for
         each dimension.}
         \end{table}
   \begin{figure}
   \centering
   \includegraphics[width=8cm]{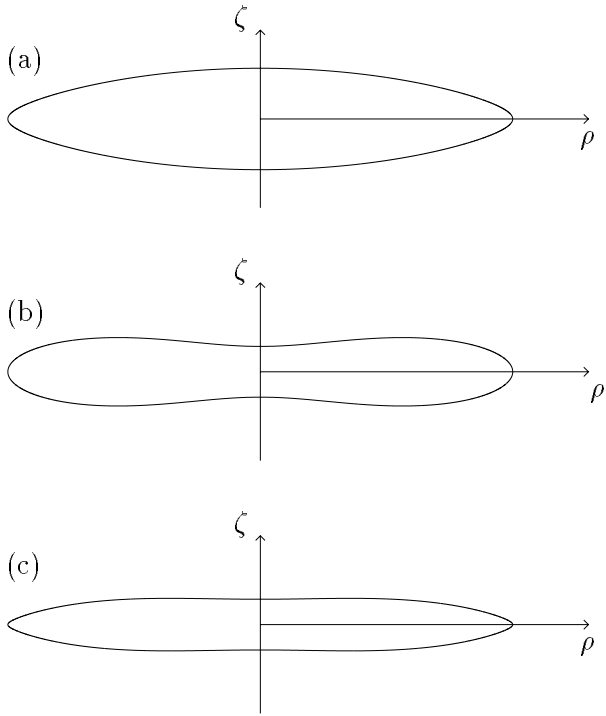}
   \caption{Meridional cross--sections of the solutions
   specified in Table 2.}
   \end{figure}
In order to check these conjectures, we have calculated
three relativistic models in the
vicinity of $R=R_1$. These models are characterized by parameters as shown in
Table 2 and their (coordinate) shapes are depicted in Fig.~1. Note that the
high accuracy of our code is crucial when investigating the subtle behaviour of
the relativistic solutions in this region --- an earlier attempt by Butterworth
(\cite{but}) could not finally clarify these questions. Solution (a)
shows indeed a `lens shape', whereas solution (b) has a
`one--ring tendency' as does solution (c), albeit far less pronounced.
The solution (a) is rather close to the Newtonian
`lens shape' sequence. The situation becomes more and more complex when the
bifurcation points $R_2=0.36633$\dots, $R_3=0.43527$\dots, etc. (corresponding to
$\epsilon_2=0.99375$\dots, $\epsilon_3=0.99657$\dots, etc.) of the Newtonian
two--ring, three--ring, etc. sequences are approached (for the two--ring
sequence see Eriguchi \& Hachisu \cite{eri2}). Our solution (c) already has a
value for $R$ close to $R_2$, see Table 2.

A detailed analysis of highly flattened, rotating Newtonian as well as
Einsteinian bodies including a discussion of stability aspects and
of the route to infinitesimally thin, relativistic
discs (Bardeen \& Wagoner \cite{bw1, bw2}, Neugebauer \& Meinel \cite{neumei1,
neumei2}) will be the subject of a future publication.


\begin{acknowledgement}
      This work was supported by the German
      \emph{Deut\-sche For\-schungs\-ge\-mein\-schaft\/} (DFG--project
      ME~1820/1).
\end{acknowledgement}

\end{document}